# Ultrafast domain wall motion in ferrimagnets induced by magnetic anisotropy gradient

W. H. Li[1], Z. Jin[1], D. L. Wen[1], X. M. Zhang[1], M. H. Qin[1,*], and J. –M. Liu[2]

[1]*Guangdong Provincial Key Laboratory of Quantum Engineering and Quantum Materials, and Institute for Advanced Materials, South China Academy of Advanced Optoelectronics, South China Normal University, Guangzhou 510006, China*

[2]*Laboratory of Solid State Microstructures and Innovative Center for Advanced Microstructures, Nanjing University, Nanjing 210093, China*

**[Abstract]** The ultrafast magnetic dynamics in compensated ferrimagnets not only provides information similar to antiferromagnetic dynamics, but more importantly opens new opportunities for future spintronic devices [Kim *et al.*, Nat. Mater. 16, 1187 (2017)]. One of the most essential issues for device design is searching for low-power-consuming and high-efficient methods of controlling domain wall. In this work, we propose to use the voltage-controlled magnetic anisotropy gradient as an excitation source to drive the domain wall motion in ferrimagnets. The ultrafast wall motion under the anisotropy gradient is predicted theoretically based on the collective coordinate theory, which is also confirmed by the atomistic micromagnetic simulations. The antiferromagnetic spin dynamics is realized at the angular momentum compensation point, and the wall shifting has a constant speed under small gradient and can be slightly accelerated under large gradient due to the broadened wall width during the motion. For nonzero net angular momentum, the Walker breakdown occurs at a critical anisotropy gradient significantly depending on the second anisotropy and interfacial Dzyaloshinkii-Moriya interaction, which is highly appreciated for further experiments including the materials selection and device geometry design. More importantly, this work unveils a low-power-consuming and high-efficient method of controlling the domain wall in ferrimagnets, benefiting to future spintronic applications.

---

Email: qinmh@scnu.edu.cn

## I. Introduction

Antiferromagnetic materials show fast magnetic dynamics and produce non-perturbing stray fields, attributing to their zero magnetization and ultralow susceptibility. These advantages make them promising candidates for next generation of high-density and high-speed spintronic devices.[1-5] However, the magnetic field immunity of antiferromagnetic materials also hinders the detection and manipulation of magnetic states.[6-8] Thus, it is still challenging to experimentally study the antiferromagnetic spin dynamics, although several stimuli have been predicted to drive the fast domain wall motion in the earlier theoretical works.[9-18] Therefore, a reliable and direct detection of the magnetic states remains to be a common issue for antiferromagnetic spintronic researches.

To overcome this deficiency, an immediate alternative strategy is to consider ferrimagnetic (FiM) systems where the fast magnetic dynamics in the vicinity of angular momentum compensation temperature $T_A$ can be achieved,[19] at which the net momentum vanishes while the net magnetic moment is nonzero. It has been theoretically predicted and experimentally confirmed that the FiM dynamics at $T_A$ is similar to the antiferromagnetic dynamics. More importantly, the magnetic states of a FiM system at $T_A$ can be effectively detected and addressed through the magnetoelectric[20-22] and magneto-optical[23] responses, benefiting from their nonzero magnetic moment, and thus highly appreciated.

In fact, the magnetic field- and electrical current-driven fast domain wall motions in angular momentum compensated ferrimagnets have been experimentally reported, respectively.[19,24-26] Also, the Walker breakdown field, under which the domain wall begins to precess and reaches to a threshold speed, is significantly increased and the domain wall mobility is extensively enhanced when the net angular momentum approaches to zero. At $T_A$, the field diverges, and the domain wall speed keeps increasing linearly with field due to the excluded Walker breakdown, exactly the same as in antiferromagnets. For example, the domain wall speed as high as ~ 20 km s$^{-1}$T$^{-1}$ was reported at $T_A$ in rare earth 3d transition metal ferrimagnets.[19] Thus, the magnetic dynamics in ferrimagnets at $T_A$ not only provides equivalent information for antiferromagnetic spin dynamics, but more importantly opens new opportunities for future spintronic devices.

On the other hand, searching for well-controlled and low-power-consumed methods to modulate FiM domain wall is one of the most important issues for spintronic device operation, noting that the shortcomings of these proposed schemes may be detrimental for future applications. For instance, the dispersion characteristic of magnetic field generally limits the density of ferrimagnetic elements and hinders the further optimization of device dimension.

Moreover, some of the electrical current related schemes normally generate Joule heating and unnecessary energy loss, significantly affecting the data transportation process where a stable operating temperature is benefiting. Along this line, electric field control could be highly preferred,[27] to be explained in detail below.

First, numerous experiments have revealed the voltage control of magnetism. For example, the voltage induced magnetic anisotropy gradient has been experimentally reported in magnetic heterostructures through elaborate structure design.[28-30] Under such a gradient, the magnetic domain wall tends to move towards the low anisotropy side in order to save free energy. As a matter of fact, the anisotropy gradient has been proven to efficiently drive the skyrmions motion and antiferromagnetic domain wall motion,[31-33] and this scheme could be also utilized to control the FiM domain wall motion. More importantly, this alternative scheme is promising for future spintronic applications considering the low-energy cost and the high operating efficiency. However, as far as we know, few works on this subject have been reported, while the dynamics of FiM domain wall under anisotropy gradient is certainly an urgent topic to be understood, in order to provide instruction for future experiments and promote the application process for spintronics.

In this work, we study the domain wall dynamics of ferrimagnets under an anisotropy gradient, using the collective coordinate theory and atomistic Landau-Lifshitz-Gilbert (LLG) simulations. It is demonstrated that the wall speed and precession direction depend closely on the net angular momentum. At the angular momentum compensation point, the Walker breakdown vanishes and the wall moves at a maximal speed, similar to the case of antiferromagnetic dynamics. It will be shown that the wall remains to shift at a constant speed under small gradient, while the motion can be slightly accelerated under large gradient due to the broadened wall width during the motion. Furthermore, for a nonzero angular momentum, the Walker breakdown gradient could be modulated by utilizing a second anisotropy and the interfacial Dzyaloshinkii-Moriya (DM) interaction. These results provide useful information for future material design and spintronic applications.

## II. Analytical analysis and numerical simulation

We investigate theoretically the domain wall motion for ferrimagnets such as rare earth and transition metal compounds, whose magnetic structure is depicted in Fig. 1(a) where the spins of two inequivalent sublattices are coupled antiferromagnetically.[34] We set $\mathbf{n}_{1,2}(\mathbf{r}, t)$ ($\mathbf{n}_1 = -\mathbf{n}_2$), $\mathbf{M}_{1,2}$ ($\mathbf{M}_{1,2} = M_{1,2} \mathbf{n}_{1,2}$), $\gamma_{1,2}$, $g_{1,2}$, and $\alpha_{1,2}$ to be the local unit vector at time $t$ and position **r**, magnetization moment, gyromagnetic ration, Landé $g$ factor, and Gilbert damping constant of the two sublattices. Thus, the spin density of the sublattice $i$ is given by $s_i = M_i/\gamma_i$ with $\gamma_i = g_i\mu_B/\hbar$, where $\mu_B$ is the Bohr magneton. It is noted that the net magnetization **M** = **M**$_1$ + **M**$_2$ is nonzero at $T_A$ where the net angular momentum $\delta_s = s_1 - s_2 = 0$, because of the different Landé $g$ factors between the two sublattices.

*2.1. Analytical treatment*

Following the collective coordinate approach, the low-temperature magnetic dynamics of FiM model is described by the Lagrangian density $L = L_B - U$ with the spin Berry phase $L_B$ and the potential-energy density $U$.[19,35] In detail, the Berry phase is associated with the staggered spin density $s = (s_1 + s_2)/2$ and the net spin density $\delta_s$, which can be described by:[18,19,35]

$$L_B = s\dot{\mathbf{n}} \cdot (\mathbf{n} \times \mathbf{m}) + \delta_s \mathbf{a}(\mathbf{n}) \cdot \dot{\mathbf{n}}, \tag{1}$$

where $\mathbf{n} \equiv (\mathbf{n}_1 - \mathbf{n}_2)/2$, and $\mathbf{m} \equiv (\mathbf{n}_1 + \mathbf{n}_2)/2$, $\dot{\mathbf{n}}$ represents the derivative with respect to time, $\mathbf{a}(\mathbf{n})$ is the vector potential generated by a magnetic monopole of unit charge satisfying $\nabla_\mathbf{n} \times \mathbf{a} = \mathbf{n}$. The potential-energy density is given by

$$U = \frac{A_{ex}}{2}(\nabla \mathbf{n})^2 + \frac{\mathbf{m}^2}{2\chi} - \frac{K(z)}{2}n_z^2 - \frac{k}{2}n_x^2 + \frac{D}{2}\mathbf{e}_y \cdot (\mathbf{n} \times \partial_z \mathbf{n}). \tag{2}$$

Here, the first and second terms are the inhomogeneous and homogeneous exchange energies where $A_{ex} > 0$ is the exchange stiffness and $\chi$ is the magnetic susceptibility. The third term is the easy-axis anisotropy along the $z$ axis (nanowire axis) with positive $K$ which changes linearly with the $z$-coordinate $K(z) = K_0 - z\, dK/dz$. The fourth term is the so-called second anisotropy or intermediate anisotropy defined along the $x$ axis with $k > 0$, and this anisotropy should be weaker than the easy-axis anisotropy along the $z$-axis. The last term is the interfacial DM interaction with $D > 0$ and $\mathbf{e}_y$ is the unit vector in the $y$ direction. To obtain an more explicit expression of the Lagrangian density, we replace **m** with $\mathbf{m} = s\chi\, \dot{\mathbf{n}} \times \mathbf{n}$,[36,37] and obtain

$$L = \frac{\rho}{2}\dot{\mathbf{n}}^2 + \delta_s \mathbf{a}(\mathbf{n}) \cdot \dot{\mathbf{n}} - \frac{A_{ex}}{2}(\nabla \mathbf{n})^2 + \frac{K}{2}n_z^2 + \frac{k}{2}n_x^2 - \frac{D}{2}\mathbf{e}_y \cdot (\mathbf{n} \times \partial_z \mathbf{n}), \tag{3}$$

where $\rho \equiv s^2\chi$ parametrizes the inertia of dynamics. The dissipative dynamics can be described by introducing the Rayleigh function density $R = s_\alpha \dot{\mathbf{n}}^2/2$ with $s_\alpha = \alpha_1 s_1 + \alpha_2 s_2$ accounting for the energy and spin loss due to the magnetic dynamics.[38]

Now we discuss the low-energy dynamics of FiM domain wall. Following the earlier work, we introduce two collective coordinates, the position $q(t)$ and azimuthal angle $\phi(t)$ in Eq. (3) to characterize the FiM domain wall under an anisotropy gradient. We consider the Walker ansatz[39] for the domain wall profile: $\mathbf{n}(z, t) = (\text{sech}((z-q)/\lambda)\cos\phi, \text{sech}((z-q)/\lambda)\sin\phi, \tanh((z-q)/\lambda))$ where $\lambda$ is the domain wall width. After applying the Euler-Lagrange equation, we obtain the equations of motion for the two coordinates:

$$M\ddot{q} + G\dot{\phi} + M\dot{q}/\tau = F, \qquad (4)$$

$$I\ddot{\phi} - G\dot{q} + I\dot{\phi}/\tau = k_0 \sin\phi\cos\phi + D_0 \sin\phi, \qquad (5)$$

where $M = 2\rho A/\lambda$ is the mass with $A$ the cross-sectional area of the domain wall, $I = 2\rho A\lambda$ is the moment of inertia, $G = 2\delta_s A$ is the gyrotropic coefficient, $\tau = \rho/s_\alpha$ is the relaxation time, $F = 4A\lambda\, dK/dz$ is the force exerted by an anisotropy gradient, $k_0 = 2k\lambda A$, and $D_0 = \pi DA/2$.

A specific solution to Eq. (4) and Eq. (5) for $k = D = 0$ gives the domain wall velocity $v$ and domain wall plane precession speed:

$$v = \frac{2\lambda^2}{\delta_s^2/s_\alpha + s_\alpha} \cdot \frac{dK}{dz}, \qquad (6)$$

$$\dot{\phi} = \frac{\delta_s}{s_\alpha \lambda} v. \qquad (7)$$

Eq. (6) shows that velocity $v$ increases linearly with $dK/dz$ and reaches the maximum at the angular momentum compensation point $T_A$ where $\delta_s$ vanishes ($\delta_s \sim 0$). To illustrate that this velocity can be high in real materials, one gives a crude estimation of $v$ by taking the well-known FiM compound GdFeCo as an example.[19,24,26] Setting the internal parameters exchange stiffness $A_{ex} = 50$ pJ/m, anisotropy constant at high anisotropy end $K_0 = 0.5$ MJ/m$^3$, $M_1 = 440$ kA/m, $M_2 = 400$ kA/m, $\alpha_1 = \alpha_2 = 0.01$, $g_1 = 2.2$, and $g_2 = 2.0$, one obtains a wall motion velocity $v \sim 1.2$ km/s at the compensation point under an anisotropy gradient $dK/dz = 300$ GJ/m$^4$, comparable to the current- and the field-driven motions for antiferromagnetic domain wall motions. Furthermore, as shown in Eq. (7), the domain wall plane rotates with the domain wall propagation without any favored orientation due to $k = 0$, which is closely dependent of $\delta_s$.

*2.2. Numerical calculation*

In order to check the validity of the above analytical treatment, we also perform the numerical simulations based on the atomistic LLG equation. Here, the corresponding one-dimensional discrete Hamiltonian is given by:[40]

$$H = J\sum_i \mathbf{S}_i \cdot \mathbf{S}_{i+1} - \sum_i K_i (S_i^z)^2 - K_x \sum_i (S_i^x)^2 + \sum_i \mathbf{D}_i \cdot (\mathbf{S}_i \times \mathbf{S}_{i+1}), \quad (8)$$

where the first term is the exchange interaction with $J = 1$, $\mathbf{S}_i$ is the normalized spin moment vector at lattice site $i$. The second term is the anisotropy energy with the easy axis along the $z$-direction, and the anisotropy constant at site $i$ is described by $K_i = K_0 - ia\,\Delta K$ where $\Delta K$ describes the anisotropy gradient magnitude, $a$ is the lattice constant. The third term is the second anisotropy $K_x$ along the $x$-axis, and the last term is the DM interaction with $\mathbf{D}_i = (0, D_y, 0)$.

Then, the dynamics is investigated by solving the stochastic LLG equation,[41-43]

$$\frac{\partial \mathbf{S}_i}{\partial t} = -\frac{\gamma_i}{M_i(1+\alpha_i)^2} \mathbf{S}_i \times [\mathbf{H}_i + \alpha_i (\mathbf{S}_i \times \mathbf{H}_i)], \quad (9)$$

where $\mathbf{H}_i = -\partial H/\partial \mathbf{S}_i$ is the effective field. Without loss of generality, we set the damping constants $\alpha_1 = \alpha_2 = 0.01$, the gyromagnetic ratios $\gamma_1 = 1.1$ and $\gamma_2 = 1.0$ corresponding to the Landé $g$-factors $g_1 = 2.2$ and $g_2 = 2.0$ for the two sublattices.[44]

To investigate the dynamics in the vicinity of the momentum compensation point, several sets of $(M_1, M_2)$ are employed, as listed in Table I. Unless stated elsewhere, the LLG simulations are performed on a $1 \times 1 \times 400$ lattices with open boundary conditions using the fourth-order Runge-Kutta method with a time step $\Delta t = 1.0 \times 10^{-4}$ $\mu_s/J\gamma_{eff}$ where $\mu_s$ is the saturation moment and $\gamma_{eff} = (\gamma_1 + \gamma_2)/2$. After a sufficient relaxation of the domain structure, the anisotropy gradient is applied to drive the domain wall motion, as schematically depicted in Fig. 1(a).

As a matter of fact, a comparison between the analytical treatment and the atomic model can be useful in qualitative sense. It is seen from the atomistic model that various torques act on the wall spins.[16] The two spins neighboring the central wall spin deviate differently from the easy-axis with $\theta_1 > \theta_2$, resulting in the net damping torque $\Gamma^d$ from the exchange interaction on the central spin, as depicted in Fig. 1(b). The damping torques $\Gamma^d \sim -\mathbf{S} \times (\mathbf{S} \times \mathbf{H})$ point in an opposite direction on the two sublattices and drive the wall motion. Moreover, the precession torques $\Gamma^p \sim -\mathbf{S} \times \mathbf{H}$ pointing into the same direction on the two sublattices are unequal in magnitude in the case of $\delta_s \neq 0$, resulting in the precession of the wall plane with the wall propagation, in agreement with Eq. (7). For $\delta_s = 0$, torques $\Gamma^p$ on the two sublattices are equal and the domain wall plane is fixed.

Thus, the fast domain wall motion and the precession of the wall plane in ferrimagnets are theoretically revealed and qualitatively confirmed by the atomic model simulations. Subsequently, we present the analytically derived and numerically calculated results to demonstrate the quantitative consistence between the analytical derivation and atomistic simulation on one hand, and more importantly to unveil the FiM dynamics in details.

## III. Results and discussion

### 3.1. Domain wall dynamics

We first present the domain wall dynamics by discussing the wall velocity and precession speed as a function of the anisotropy gradient respectively. Fig. 2(a) shows the numerically simulated (empty points) and Eq. (6)-based calculated (solid lines) wall velocity $v$ as a function of $\Delta K$ for various $\delta_s$ and $K_0 = 0.01J$, $K_x = 0$ and $D_y = 0$. It is seen that the simulated data fit the calculations perfectly, confirming the validity of the analytical treatment. Here, two issues deserve highlighting. First, the driving torque increases with the increasing $\Delta K$, which significantly enhances the wall motion speed. Specifically, $v$ increases linearly with $\Delta K$, noting that here only low anisotropy gradient is considered and the domain wall width is hardly changed during the motion. Second, for a fixed $\Delta K$, $v$ increases with decreasing magnitude of $\delta_s$, and reaches to the maximum at the angular momentum compensation point $\delta_s = 0$, as clearly shown in Fig. 2(b) where $v(\delta_s)$ curves for various $\Delta K$ are presented.

We then discuss the domain wall plane precession which appears for a nonzero $\delta_s$ in accompanying with the wall motion, as shown in Fig. 2(c) where the angular velocity ($d\phi/dt$) of the plane as a function of $\Delta K$ is plotted. Also two issues are highlighted. First, the angular velocity increases linearly with $\Delta K$ or the wall speed $v$. In comparison with the dynamics for

$\delta_s = 0$ where the domain wall plane is fixed, the wall plane precession leads to additional energy dissipation, resulting in the low wall mobility for nonzero $\delta_s$ under the same $\Delta K$. Second and more interestingly, the precession direction of the wall plane depends on the sign of $\delta_s$. Specifically, the wall plane precesses clockwise around the easy-axis for $\delta_s > 0$, while does counterclockwise for $\delta_s < 0$, as clearly shown in Fig. 2(d) where the *x* and *y* components of local quantity **n**, $n_x$ and $n_y$, are presented at various times for $\delta_s > 0$ (top half) and $\delta_s < 0$ (bottom half). With the wall motion, opposite precession directions are clearly observed in the two cases, in consistent with the theoretical prediction in Eq. (7).

So far, the anisotropy gradient driven domain wall motion in the vicinity of the angular momentum compensation point of ferrimagnets has been clearly uncovered in our theoretical analysis and LLG simulations. In experiments, anisotropy gradient could be induced by tuning electric field on particular heterostructures, and efficiently drives the domain wall motion generating Joule heat much less than those electrical current related methods. Thus, the proposed method in the work is expected to be both low power-consuming and high-efficient, which is essential for future spintronic applications.

*3.2. Roles of internal parameters*

Based on the good consistency between the analytical analysis and numerical calculations, one is able to discuss the roles of various internal parameters. An unveiling of these roles would be highly appreciated for practical applications including the materials selection, device geometry design, and performance optimization.

First, the anisotropy constant $K_0$ determines the wall width $\lambda$ which can be estimated by $a(J/2K_0)^{1/2}$, and in turn affects the wall speed which increases with $\lambda$ as demonstrated in Eq. (6). Thus, contrary to $\Delta K$, a large $K_0$ results in a small $\lambda$ and makes the wall motion slow, as confirmed in our simulations. In Fig. 3(a), the simulated and calculated speeds as functions of $K_0$ for various $\Delta K$ at the angular momentum compensation point are plotted, not only showing the excellent consistence between the simulation and analytical derivation but also clearly revealing that the anisotropy magnitude enables an decelerated wall motion. In addition, an enhanced damping term always reduces the wall mobility,[45,46] and *v* decreases with the increase of the damping constant *α*. As a matter of fact, *v* linearly increases with $1/\alpha$, as shown in Fig. 3(b) where presents the simulated and calculated $v$ as functions of $1/\alpha$ at $\delta_s = 0$ for fixed $K_0$ and $\Delta K$.

In the above analysis, the wall width $\lambda$ is simply set to be unchanged during the wall motion, which well describes the case of very low anisotropy gradient. However, when the wall shifts under a high gradient, the wall is considerably enlarged, resulting in an accelerated domain wall motion. This phenomenon has been also observed in our simulations (dashed lines) and calculations (solid lines) in Fig. 4 which give the evolution of the wall position (Fig. 4(a)) and the local wall velocity (Fig. 4(b)) for various $\Delta K$ at $\delta_s = 0$. In this case, the wall width could be updated to $\lambda = a(J/2K_c)^{1/2}$ with $K_c$ the anisotropy on the wall central spin. A constant velocity is obtained under a low gradient $\Delta K \sim 0.5 \times 10^{-5} J/a$, while an acceleration of the wall motion under a high gradient $\Delta K \sim 2 \times 10^{-5} J/a$ is clearly observed.

Second, the intermediate anisotropy $K_x$ could be non-negligible in some FiM materials, and affects the wall motion. Subsequently, we check the effect of $K_x$ on the dynamics of domain wall in ferrimagnets. The time evolutions of the wall position for various $K_x$ for $\delta_s = 0.022$ are presented in Fig. 5(a) which exhibits three types of wall motion.[16] As discussed above, the wall has no favored orientation for $K_x = 0$ and rotates continuously and moves constantly with a reduced speed. The consideration of the intermediate anisotropy suppresses the rotation of the wall plane, and in turn significantly affects the wall motion. In the case of small anisotropy of $K_x = 2.5 \times 10^{-5} J$, the Walker breakdown occurs under the anisotropy gradient $\Delta K$ larger than the threshold value $\Delta K_{WB}$. Here, the Walker breakdown gradient $\Delta K_{WB}$ can be estimated by $K_x s\alpha/4\delta_s\lambda$.[47,48] In the case of high anisotropy $K_x = 10 \times 10^{-5} J$, the precession of the domain wall is completely suppressed for the considered $\Delta K$, resulting in the wall motion with a maximal velocity. On the other hand, the wall motion at the angular momentum compensation point $\delta_s = 0$ where no precession of the wall is available is independent of the anisotropy $K_x$, as clearly shown in Fig. 5(b) where presents the mean velocity of the domain wall as a function of $K_x$ for $\delta_s = 0$ and $\delta_s = 0.022$. Moreover, for a fixed gradient $\Delta K$ below the Walker breakdown $\Delta K_{WB}$, $\delta_s = 0.022$ is with a magnetization smaller than $\delta_s = 0$, and the domain wall motion for $\delta_s = 0.022$ is slightly faster than $\delta_s = 0$.

Third, a DM interaction could be induced at interface between heavy metal and ferrimagnet and modulated efficiently through elaborate heterostructure design. Similarly, the interfacial DM interaction $D_y(0, 1, 0)$ also suppresses the precession of the wall plane and speeds up the wall motion below $\Delta K_{WB}$. In Fig. 5(c), the simulated velocities as a function of $D_y$ for $\delta_s = 0$ and $\delta_s = 0.022$ for $\Delta K = 0.5 \times 10^{-5} J/a$ is plotted, revealing the critical DM interactions $D_c$ which can be given by $|D_c| = 8\delta_s\lambda^2 \Delta K_{WB} / \pi s\alpha$ for nonzero $\delta_s$.[26,47,48] Under a fixed $\Delta K$, the Walker breakdown occurs for $|D_y| < |D_c|$, while vanishes for $|D_y| > |D_c|$. The simulated $|D_c|$ (empty points) as a function of $\delta_s$ for various $\Delta K$ is presented in Fig. 5(d), well in consistent with the theoretical derivation (solid lines). Thus, this prediction could be used to improve the Walker breakdown field and to enhance the domain wall mobility, which is very meaningful for spintronic applications.

Thus, the domain wall motion depending on the internal parameters has been clearly unveiled, which definitely provides useful information for future material selection and device design. For example, high domain wall mobility could be available in ferrimagnet with not too large $K_0$ and considerable second anisotropy, as suggested in our calculations. Moreover, a large DM interaction generated in interface between heavy metal and ferrimagnet significantly improves the Walker breakdown field and ensures the fast domain wall motion. Of cause, these predictions given here deserve to be checked in further experiments.

## IV. Conclusion

To summarize, we have studied analytically and numerically the dynamics of the domain wall in ferrimagnets driven by the magnetic anisotropy gradient. The wall moves towards the low anisotropy side to release the free energy and reaches to a maximal velocity at the angular momentum compensation point where exhibits the antiferromagnetic dynamics. Moreover, the net spin angular momentum determines not only the wall velocity but also the precession direction of the domain wall plane. Furthermore, for nonzero net angular momentum, Walker breakdown occurs under a critical anisotropy gradient which significantly depends on the intermediate anisotropy and interfacial DM interaction. This work unveils a low power-consuming and also high-efficient method of controlling the domain wall in ferrimagnets, benefiting to future experiments design and spintronic applications.

**Acknowledgment**

The work is supported by the National Key Projects for Basic Research of China (Grant No. 2015CB921202), and the Natural Science Foundation of China (No. 51971096), and the Science and Technology Planning Project of Guangzhou in China (Grant No. 201904010019), and the Natural Science Foundation of Guangdong Province (Grant No. 2016A030308019).

*References:*

1. T. Jungwirth, X. Marti, P. Wadley and J. Wunderlich, Nat. Nanotechnol. **11**, 231 (2016).
2. P. Wadley *et al*., Science **351**, 587 (2016).
3. J. Železný, P. Wadley, K. Olejník, A. Hoffmann and H. Ohno, Nat. Phys. **14**, 220 (2018).
4. J. Torrejon *et al*., Nature **547**, 428 (2017).
5. P. Park *et al*., npj Quantum Mater. **3**, 63 (2018).
6. O. Gomonay, T. Jungwirth, and J. Sinova, Phys. Rev. Lett. **117**, 017202 (2016).
7. T. Shiino, S. H. Oh, P. M. Haney, S. W. Lee, G. Go, B. G. Park, and K. J. Lee, Phys. Rev. Lett. **117**, 087203 (2016).
8. V. G. Barya khtar, B. A. Ivanov, and M. V. Chetkin, Sov. Phys. Usp. **28**, 563 (1985).
9. K. M. D. Hals, Y. Tserkovnyak, and A. Brataas, Phys. Rev. Lett. **106**, 107206 (2011).
10. A. Qaiumzadeh, L. A. Kristiansen, and A. Brataas, Phys. Rev. B **97**, 020402(R) (2018).
11. T. Shiino, S.-H. Oh, P. M. Haney, S.-W. Lee, G. Go, B.-G. Park, and K.-J. Lee, Phys. Rev. Lett. **117**, 087203 (2016).
12. O. Gomonay, T. Jungwirth, and J. Sinova, Phys. Rev. Lett. **117**, 017202 (2016).
13. E. G. Tveten, A. Qaiumzadeh, and A. Brataas, Phys. Rev. Lett. **112**, 147204 (2014).
14. S. K. Kim, Y. Tserkovnyak, and O. Tchernyshyov, Phys. Rev. B **90**, 104406 (2014).
15. S. K. Kim, O. Tchernyshyov, and Y. Tserkovnyak, Phys. Rev. B **92**, 020402(R) (2015).
16. S. Selzer, U. Atxitia, U. Ritzmann, D. Hinzke, and U. Nowak, Phys. Rev. Lett. **117**, 107201 (2016).
17. Z. R. Yan, Z. Y. Chen, M. H. Qin, X. B. Lu, X. S. Gao, and J.-M. Liu, Phys. Rev. B **97**, 054308 (2018).
18. E. G. Tveten, T. Muller, J. Linder, and A. Brataas, Phys. Rev. B **93**, 104408 (2016).
19. K.-J. Kim *et al*., Nat. Mater. **16**, 1187 (2017).
20. J. Finley and L. Q. Liu, Phys. Rev. Appl. **6**, 054001 (2016).
21. R. Mishra, J. W. Yu, X. P. Qiu, M. Motapothula, T. Venkatesan, and H. Yang, Phys. Rev. Lett. **118**, 167201 (2017).
22. N. Roschewsky, T. Matsumura, S. Cheema, F. Hellman, T. Kato, S. Iwata, and S. Salahuddin, Appl. Phys. Lett. **109**, 112403 (2016).

23. I. Radu *et al.*, Nature (London) **472**, 205 (2011).

24. S. A. Siddiqui, J. H. Han, J. T. Finley, C. A. Ross, and L. Q. Liu, Phys. Rev. Lett. **121**, 057701 (2018).

25. L. Caretta *et al.*, Nat. Nanotechnol. **13**, 1154–1160 (2018).

26. S.-H. Oh, S. K. Kim, D.-K. Lee, G. Go, K.-J. Kim, T. Ono, Y. Tserkovnyak, and K.-J. Lee, Phys. Rev. B **96**, 100407(R) (2017).

27. K. M. Cai et *al.*, Nat. Mater. **16**, 712–716 (2017).

28. R. Tomasello, S. Komineas, G. Siracusano, M. Carpentieri, and G. Finocchio, Phys. Rev. B **98**, 024421 (2018).

29. H. Xia, C. Song, C. Jin, J. Wang, J. Wang, and Q. Liu, J. Magn. Magn. Mater. **458**, 57 (2018).

30. C. Ma, X. Zhang, J. Xia, M. Ezawa, W. Jiang, T. Ono, S. N. Piramanayagam, A. Morisako, Y. Zhou, and X. Liu, Nano Lett. **19**, 353 (2019).

31. C. Ching, I. Ang, W. L. Gan and W. S. Lew, New J. Phys. **21**, 043006 (2019).

32. L. C. Shen, J. Xia, G. P. Zhao, X. C. Zhang, M. Ezawa, O. A. Tretiakov, X. X. Liu, and Y. Zhou, Phys. Rev. B **98**, 134448 (2018).

33. D. L. Wen, Z. Y. Chen, W. H. Li, M. H. Qin, D. Y. Chen, Z. Fan, M. Zeng, X. B. Lu, X. S. Gao, and J.-M. Liu, arXiv:1905.06695 (2019).

34. O. A. Tretiakov, D. Clarke, G.-W. Chern, Y. B. Bazaliy, and O. Tchernyshyov, Phys. Rev. Lett. **100**, 127204 (2008).

35. S. K. Kim, K.-J. Lee, and Y. Tserkovnyak, Phys. Rev. B **95**, 140404(R) (2017).

36. V. S. Gerasimchuk and A. A. Shitov, Low Temp. Phys. **27**, 125 (2001).

37. B. A. Ivanov and A. L. Sukstanski, Sov. Phys. JETP **57**, 214 (1983).

38. H. Goldstein, C. Poole, and J. Safko, Classical Mechanics, 3rd ed. (Addison Wesley, 2002).

39. L. D. Landau and E. M. Lifshitz, *Electrodynamics of Continuous Media*, Course of Theoretical Physics Vol. 8 (Pergamon, Oxford, 1960).

40. F. D. M. Haldane, Phys. Rev. Lett. **50**, 1153 (1983).

41. N. Kazantseva, U. Nowak, R. W. Chantrell, J. Hohlfeld, and A. Rebei, Europhys. Lett. **81**, 27004 (2007).

42. D. Landau and E. Lifshitz, Phys. Z. Sowjetunion **8**, 153 (1935).
43. T. L. Gilbert, IEEE Trans. Magn. **40**, 3443 (2004).
44. J. Jensen and A. R. Mackintosh, *Rare earth magnetism* (Clarendon Oxford, 1991).
45. N. L. Schryer and L. R. Walker, J. Appl. Phys. **45**, 5406 (1974).
46. S. Moretti, M. Voto, and E. Martinez, Phys. Rev. B **96**, 054433 (2017).
47. Z. Y. Chen, M. H. Qin, and J.-M. Liu, Phys. Rev. B **100**, 020402(R) (2019).
48. A. Mougin, M. Cormier, J. Adam, P. Metaxas, and J. Ferr, Europhys. Lett. **78**, 57007 (2007).

**Table I. Parameters chosen for the simulation.**

| Parameter | 1 | 2 | 3 | 4 | 5 | 6 | 7 |
|---|---|---|---|---|---|---|---|
| $M_1$ | 1.13 | 1.12 | 1.11 | 1.1 | 1.09 | 1.08 | 1.07 |
| $M_2$ | 1.0 | 1.04 | 1.02 | 1.0 | 0.98 | 0.96 | 0.94 |
| $\delta_s$ | -0.03273 | -0.0218 | -0.0109 | 0 | 0.0109 | 0.0218 | 0.03273 |

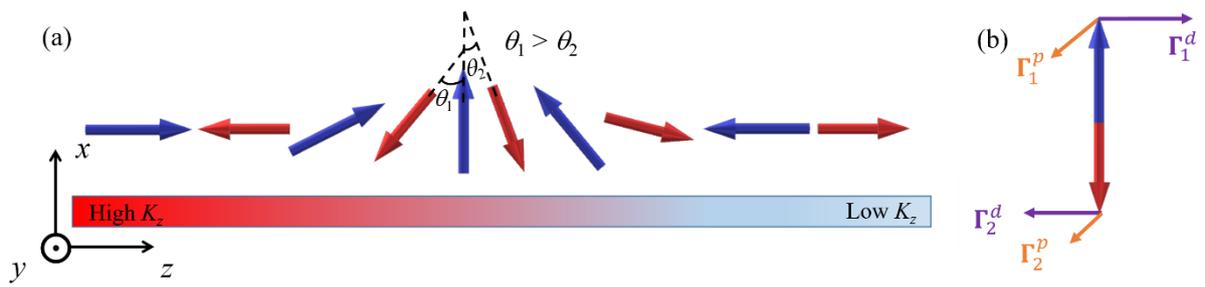

FIG.1. (color online) (a) Illustration of a domain wall in ferrimagnetic nanowire under an anisotropy gradient. Here the asymmetry of the domain wall center is exaggerated. (b) A schematic depiction of torques acting on the central spins of the domain wall.

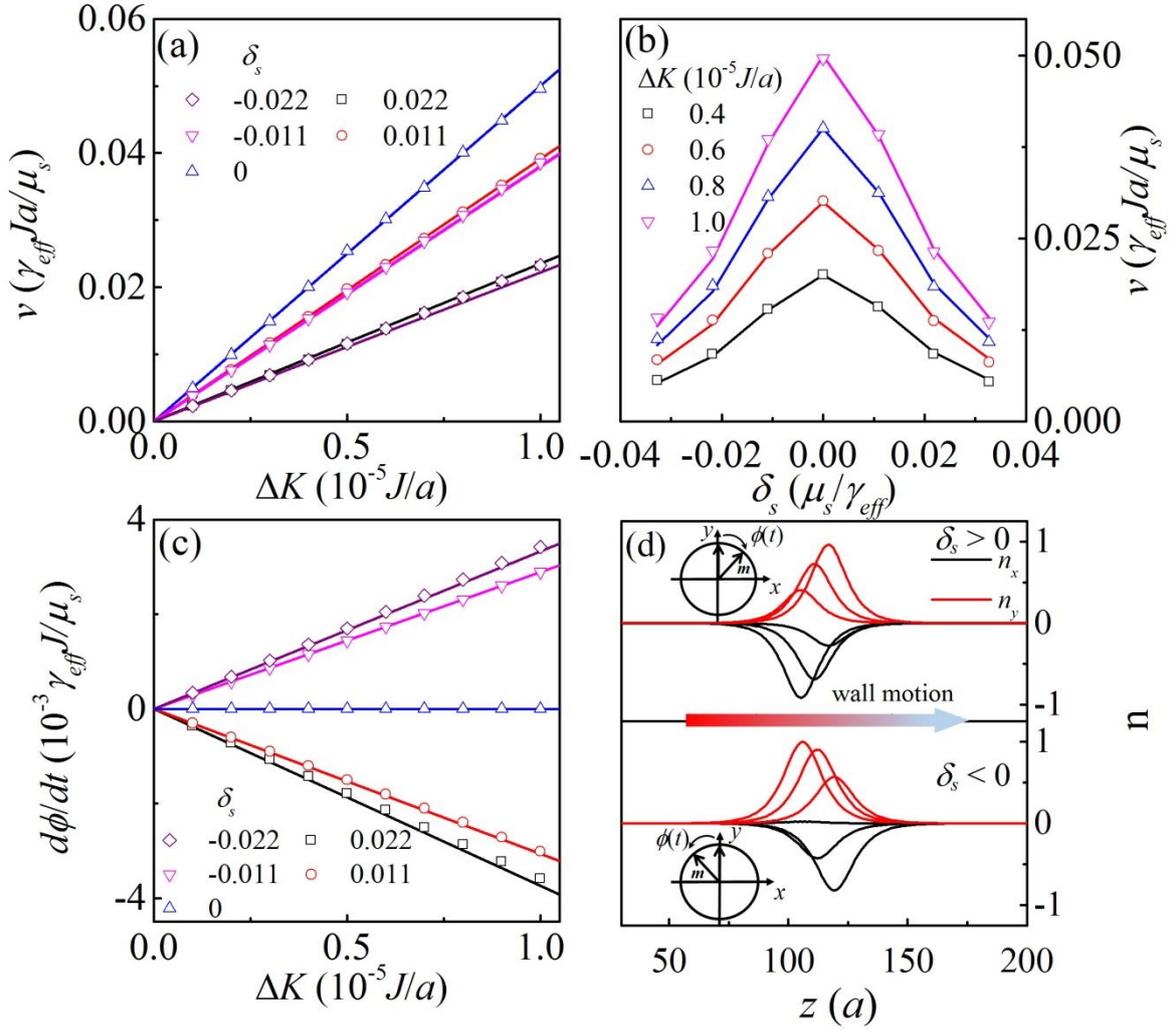

FIG.2. (color online) The simulated (empty points) and calculated (solid lines) velocities as functions of (a) $\Delta K$ for various $\delta_s$, and (b) $\delta_s$ for various $\Delta K$ for $K_0 = 0.01$. (c) The simulated (empty points) and calculated (solid lines) angular velocities of the wall plane as functions of $\Delta K$ for various $\delta_s$, and (d) the evolutions of the local $n_x$ and $n_y$ for $\delta_s > 0$ (top half) and $\delta_s < 0$ (bottom half). The rotations of the wall plane are shown in the insert of (d).

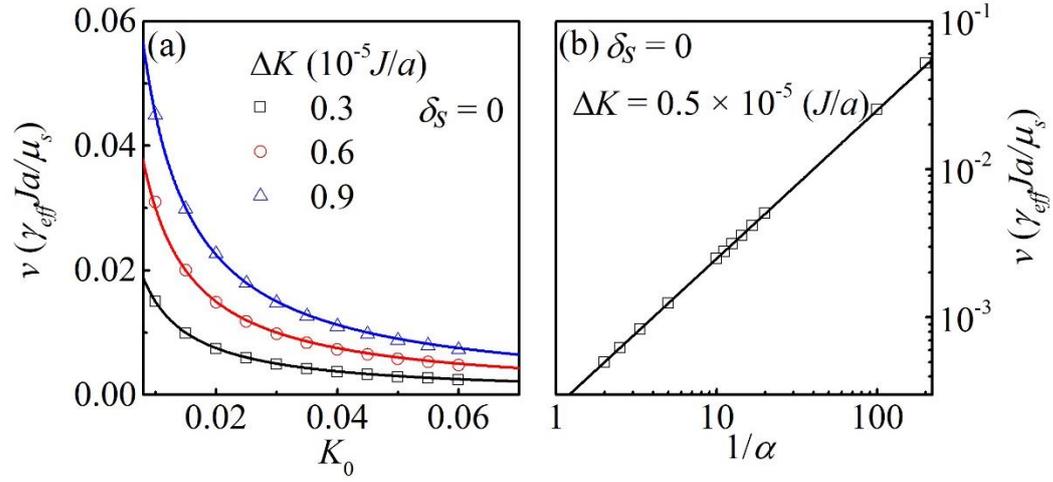

FIG.3. (color online) The simulated (empty points) and calculated (solid lines) velocities at the momentum compensation point as functions of (a) $K_0$ for various value of $\Delta K$ for $\alpha = 0.01$, and (b) $1/\alpha$ for $K_0 = 0.01J$ and $\Delta K = 0.5 \times 10^{-5}$ $J/a$.

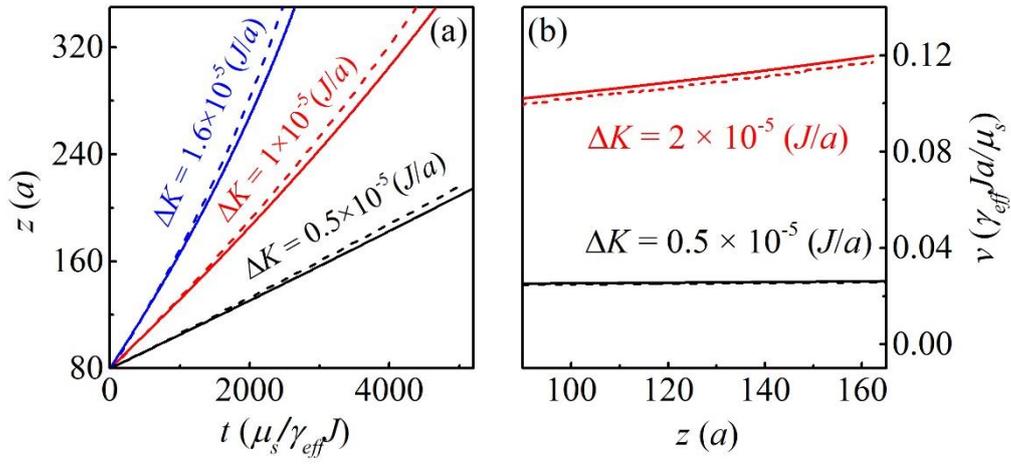

FIG.4. (color online) The simulated (dashed lines) and calculated (solid lines) (a) evolutions of the wall positions for various $\Delta K$, and (b) instantaneous speed for various $\Delta K$.

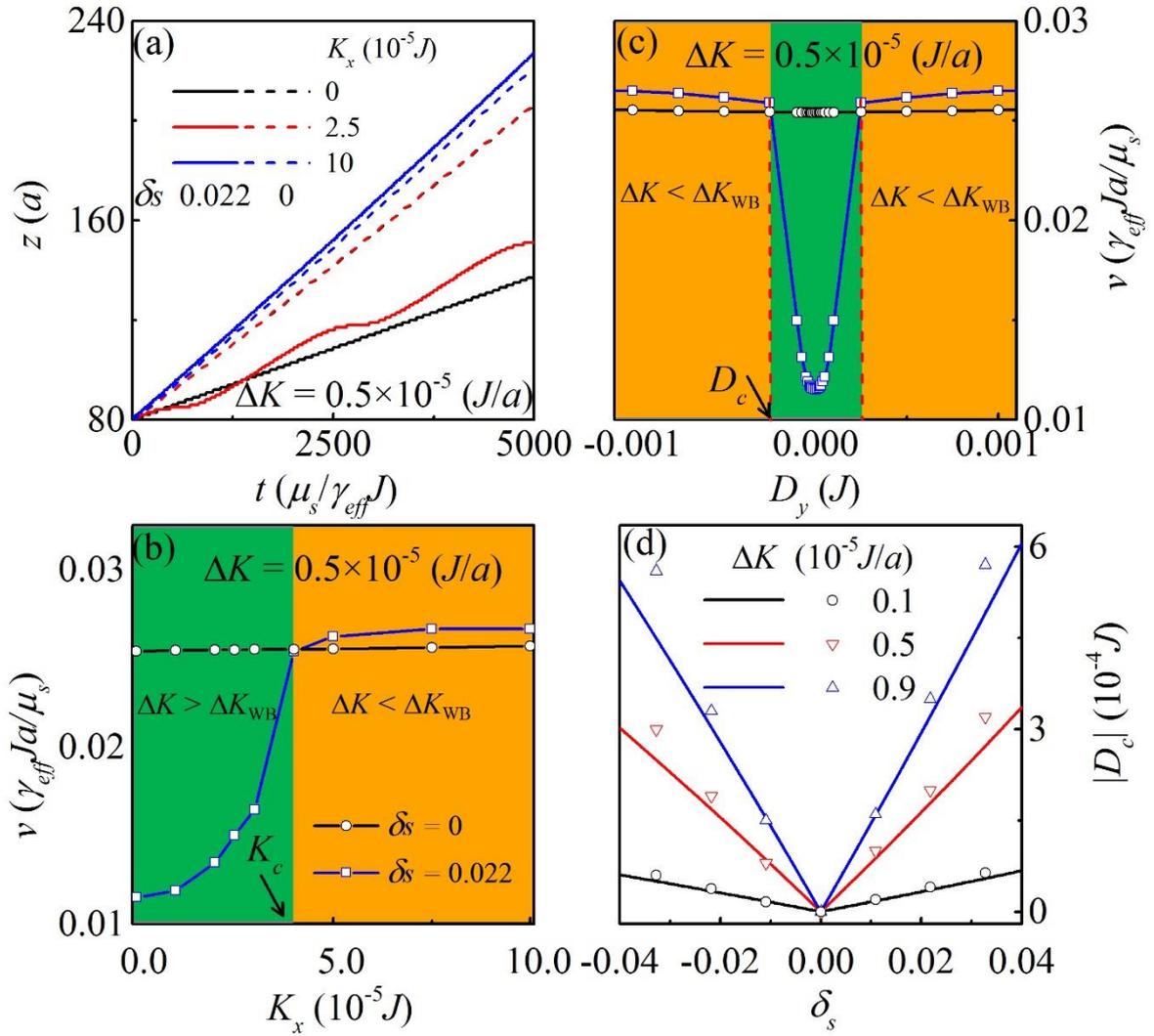

FIG.5. (color online) For $\Delta K = 0.5 \times 10^{-5}$ $J/a$, the simulated (a) evolutions of the wall position for various $K_x$, and mean velocities as functions of (b) $K_x$ and (c) $D_y$ for $\delta_s = 0$ and $\delta_s = 0.022$. (d) The simulated (empty points) and calculated (solid lines) $|D_c|$ as a function of $\delta_s$ for various $\Delta K$.